\documentclass[aps,prb,preprint,groupedaddress,showpacs,floatfix]{revtex4-1}
\usepackage{amsmath,amssymb,latexsym,epic,eepic,epsfig,graphicx,psfrag}
\usepackage[latin1]{inputenc}

\begin{document}

\title{First principles study of pentacene on Au(111)}
\author{Kurt Stokbro}
\email{kurt.stokbro@quantumwise.com}
\author{S\o ren Smidstrup}
\affiliation{QuantumWise A/S,\\
Lers{\o} Parkall\'{e} 107, DK-2100 Copenhagen, Denmark}
\homepage{http://quantumwise.com}
\date{\today}

\begin{abstract}
We investigate the atomic and electronic structure of a single layer of
pentacene on the Au(111) surface using density functional theory. To find
the candidate structures we strain match the pentacene crystal geometry with
the Au(111) surface, in this way we find pentacene overlayer structures with
a low strain. We show that the geometries obtained with this approach has
lower energy than previous proposed surface geometries of pentacene on
Au(111). We also show that the geometry and workfunction of the obtained
structures are in excellent agreement with experimental data.
\end{abstract}

\maketitle

\section{\label{sec:intro}Introduction}

The pentacene crystal(PC) is an important organic electronic material, due
to its high hole mobility, and the gold-pentacene interface is one of the
most well studied metal-organic interfaces both theoretically\cite{Ortega2011,Li-2009,Lee-2007,Lee-2005,Toyoda-2010,Saranya-2012} and
experimentally \cite{Koch-2007,McDonald-2006,Schroeder-2002,Ihm-2006,France-2003,Kafer-2007,
Watkins-2002,Kang-2003,Soe-2009,Diao-2007,Liu-2010}.

It is important to understand the geometry of the pentacene-gold interface,
since the interface properties are often close linked to geometric features.
However, previous theoretical studies of the interface geometry\cite{Li-2009, Lee-2007, Toyoda-2010} have not shown a systematic approach for
determining the lattice parameters of the pentacene overlayer. In this paper
we suggest a new approach for determining the pentacene overlayer lattice
structure, by strain matching the pentacene crystal with the gold surface.
We show that this method will recover the experimentally observed surface
geometries and we find that pentacene has a higher adsorption energy in
these geometries compared with previously suggested structures.

The paper is organised in the following way: In section~\ref{sec:methodology}
we describe the computational method, and the accuracy of the method is
verified for isolated pentacene molecules and the Au(111) surface. In
section~\ref{sec:pentacene-gold} we study the structure of a single layer of
pentacene on different Au(111) supercells and compare with previous
theoretical studies and experimental data. Finally, in section~\ref{sec:conclusions} we conclude.

\section{Methodology}

\label{sec:methodology} For the calculations we have used the Atomistix
ToolKit (ATK)~\cite{ATK12.8}, which is a density-functional theory code
using numerical localized atomic basis sets and norm conserving pseudo
potentials. For the exchange-correlation potential we have used the
Generalized Gradient Approximation (GGA) of Wang and Perdew\cite{pw91}
(PW91) as suggested by Li et. al\cite{Li-2009}.

The electronic structure is expanded in basis sets optimized to reproduce
hydrogen and carbon dimer total energies following the procedure of Blum et.
al\cite{Blum-2009}. For carbon we use 21 orbitals per atom with s, p and d
character and ranges up to 3.9 Å, while for hydrogen we use 5 orbitals per
atom with s and p character and ranges up to 4.2 Å. The size of the basis
set was chosen to converge the ionization energy of a single pentacene
molecule as described in section~\ref{sec:pentacene}. Lee \textit{et. al.}
has shown that such long range basis sets can accurately describe the weak
gold-pentacene interaction if Basis Set Superposition Errors (BSSE) are
accounted for\cite{Lee-2007}. In this study we use the counter poise
correction for the BSSE between the gold surface and the pentacene
overlayer. In the next section we will show that the computational model
gives a good description of the energetics of  pentacene molecules.

\subsection{The pentacene molecule and crystal}

\label{sec:pentacene} The first column in Table~\ref{pentacene-molecule}
shows the calculated ionisation energy of a single pentacene molecule(P1).
The ionisation energy was calculated by subtracting total energies of the
neutral and charged molecule. Both the neutral and the charged system were
relaxed until forces were less than 0.01 eV/Å. The calculations were
performed using a computational cell with multipole boundary conditions to
properly describe the long range tails of the electro-static potential. The
calculated value is 0.25 eV below the experimental value, and we relate this
discrepancy to the PW91 exchange-correlation potential.

\begin{table}
\caption{\label{pentacene-molecule}
Ionisation energy ($E_I$) and binding energy per molecule ($E_c$) for
a single pentacene (P1), dimer pentacene
(P2), and a  pentacene crystal (PC) with lattice parameters
from Ref.~\onlinecite{Sciefer-2006}. The experimental ionisation
energy of P1 is given in parentheses.
}
\begin{ruledtabular}
\begin{tabular}{cccc}
  & P1 & P2 & PC  \\ \hline
      \cline{1-4}
$E_c$ (eV) & 0.0  & 0.11  & 0.55  \\
$E_I$ (eV) & 6.34  (6.589\cite{Gruhn-2002}) & 5.82  & 5.03
\end{tabular}
\end{ruledtabular}
\end{table}

Next, we study the interaction energy of 2 pentacene molecules (P2). In this
case, we correct the total energy and forces for BSSE by using the
counterpoise correction for the two pentacene units\cite{Kohanoff-2006}.
Accouting for BSSE, we relax the system to 0.01 eV/Å\ and find an ionisation
energy of the dimer having value of 5.82 eV. Therefore, there is reduction
of the charging energy by 0.52 eV, since now the positive charge can spread
over a pair of molecules.

We calculate the binding energy between the two neutral pentacene molecules
to be 0.1~eV/pentacene. Since the PW91 functional does not properly account
for Van der Waals interactions the calculated binding energy will be too
low. Pratontep \textit{et. al.}\cite{Pratontep-2005} found the dimer binding
energy to be ~0.3 eV/pentacene using an Universal Force Field (UFF), in good
agreement with experimental data\cite{PentaceneExp}.

\begin{figure}[tbp]
\includegraphics[width=\linewidth]{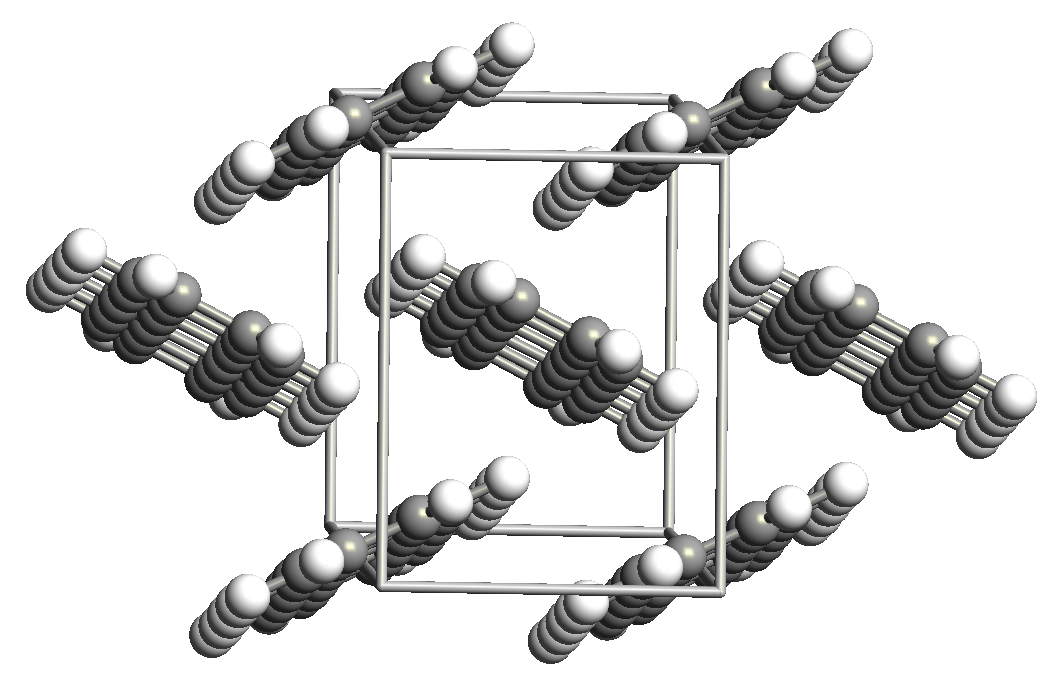}

\caption{ The figure shows the
    central pentacene molecule and the 6 surrounding molecules in the
    relaxed PC. There are 2 pentacene molecules in the
    unit cell. }  \label{fig:crystal}
\end{figure}

Finally we have set up a PC according to the crystallographic data by
Sciefer \textit{et. al.}\cite{Sciefer-2006}, (P-1: a=5.985 Å, b=7.596 Å,
c=15.6096 Å, $\alpha$=81.25$^0$, $\beta$=86.56$^0$, $\gamma$=89.8$^0$). The
atomic positions in the unit cell were relaxed using periodic boundary
conditions, still applying the BSSE correction to the pentacene units. We
found a stress on the unit cell less than 0.003 eV/Å$^3$, and the relaxation
of the lattice parameters therefore negligible. The relaxed geometry is
illustrated in Fig.~\ref{fig:crystal}.

The binding energy per pentacene molecule in the crystal is 0.55 eV, roughly
6 times larger than for the pentacene dimer, in accordance with that the
molecules in the PC are surrounded by 6 neighbours. Including Van der Waals
interactions will increase this energy substantially.

We also calculated the ionization energy of the PC by performing  charged
calculations, in this case the charge is neutralized by an uniform
background charge. The calculated ionization energy is 5.03 eV, thus a
reduction of 1.3 eV compared with the isolated molecule.

\subsection{The Au(111) surface}

To simulate gold we use a s,p,d basis set of ranges 2.7-3.6 Å, with a total
of 9 orbitals per atom, all other parameters are similar to the pentacene
calculations. With this approach we find the lattice constant of gold 4.17 Å%
\ in agreement with Li \textit{et. al.}\cite{Li-2009} which also used the
PW91 functional (experimental value 4.08 Å).

To test the description of the Au(111) surface we follow Li \textit{et. al.}%
\cite{Li-2009} and set up a Au(111)-$(\sqrt{3}\times6)$ slab consisting of 5
layers with a 15 Å\ vacuum region above the top layer. Above the top layer
we add a layer of gold ghost orbitals to get a better description of the
surface. We apply periodic boundary conditions parallel with the surface
while we use Dirichlet boundary conditions above the surface and Neumann
boundary conditions below the surface. In this way we can properly describe
different electrostatic dipoles at the two surfaces. We use a $8 \times 3
\times 1$ Monkhorst-Pack k-point grid as in Li \textit{et. al.}\cite{Li-2009}
and a Fermi-Dirac occupation scheme with room temperature broadening.

The first step is to relax the two upper layers of the gold surface until
forces are below 0.01 eV/Å. After relaxation we find a work function of gold
of 5.19 eV, slightly below the value of 5.25 eV obtained by Li \textit{et.
al.}\cite{Li-2009}.

\section{Pentacene on gold}

\label{sec:pentacene-gold}

In this section we will investigate the adsorption of pentacene (P1)  on the
Au(111) surface. Other studies suggest that the interaction between
pentacene and the gold surface is weak\cite{Li-2009}, thus, the pentacene
intermolecular interaction in the adsorption layer must therefore give an
important contribution to the adsorption energy. We will assume that this
interaction is maximum when pentacene is in its crystal geometry. Our
approach to find the geometry of the pentacene adsorption layer is therefore
to find a common supercell of the Au(111) surface and the PC. For the PC we
will assume that it is adsorbed with the phenyl rings laying flat on the
surface, corresponding to an A-C surface plane.

\subsection{Strain matching the surface cells}

In the following we discuss our algorithm for finding a common supercell of
the Au(111) and PC(010) surfaces. We first generate all possible bravais
lattices for the two surfaces,
\begin{eqnarray}
(\vec{v}_1, \vec{v }_2) & = & N (\vec{a}_1, \vec{a}_2), \\
(\vec{u}_1, \vec{u }_2) & = & M (\vec{b}_1, \vec{b}_2),
\end{eqnarray}
where $\vec{a}_1, \vec{a}_2$ are the primitive vectors for the Au(111)
surface, $N$ a $2\times2$ repetition matrix where the entries are integers
below a maximum value $N_{max}$, and $\vec{v}_1, \vec{v }_2$ the bravais
lattice vectors of the supercell. $\vec{b}_1, \vec{b}_2, M, \vec{u}_1, \vec{%
u }_2 $ are the corresponding quantities for the PC(010) surface.

We next determine a rotation matrix, $R$ which rotates $\vec{u}_1$ over in $\vec{v}_1$, and a strain tensor, $\varepsilon$ which maps the rotated
PC(010) surface lattice onto the Au(111) lattice
\begin{equation}
(\vec{v}_1, \vec{v }_2) = (1+\varepsilon) R (\vec{u}_1, \vec{u }_2).
\end{equation}

\begin{figure}[tbp]
\includegraphics[width=\linewidth]{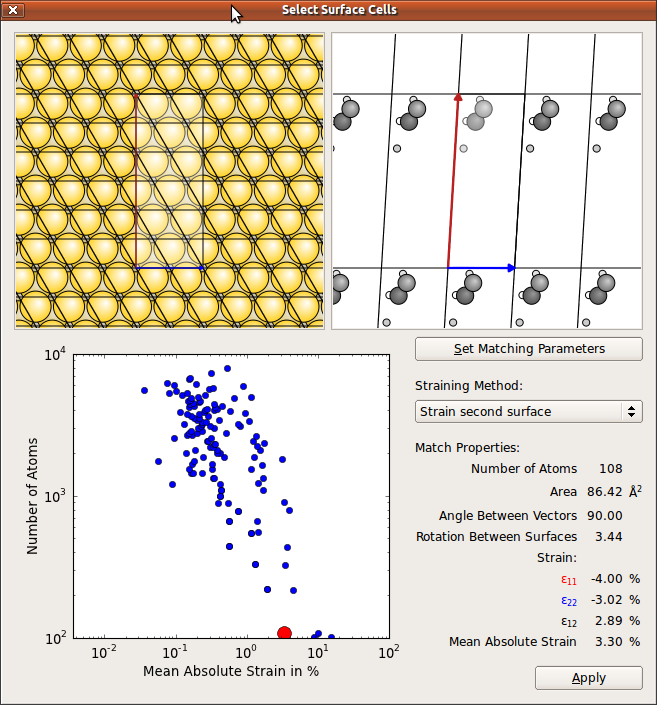}

\caption{ The graph shows the number of atoms and mean strain for
    matching a selected sets of Au(111) and PC(010) surface
    cells. The red dot in the graph corresponds to matching of the
    Au(111)-$(2\times 3\sqrt{3} )$ and the PC(010)-$(1\times1)$
    cell, which has a mean strain of 1.83\%.}  \label{fig:matching}
\end{figure}

To match the 2 crystals we generate all surface lattices with $N_{max},
M_{max} < 10$, and for each combination we calculate the strain tensor. Fig.~\ref{fig:matching} shows the strain of the different combinations as a
function of the number of atoms in the unit cell of the combined system.
Note that for both the gold and PC we have used the experimental lattice
constant. The figure shows that by increasing the number of atoms in the
surface cell the strain between the 2 lattices can be reduced.

Table~\ref{tab:strain} summarizes the lattices with the lowest strain for a
low repetition of the PC(010) surface lattice.  The table also include the
Au(111)-$(\sqrt{3}\times 6 )$ lattice, as seen from the table the strain is
rather large for this structure. The $(\sqrt{3}\times6)$ structure was
investigated by Li \textit{et. al.}\cite{Li-2009} using the VASP\cite{vasp}
code and by comparing with their results we can check the accuracy of our
computational approach.

\begin{table}
\caption{\label{tab:strain}Strain in  the PC(010) lattice to match
 different Au(111) super cells. The first columns shows the length of
 the Au(111) surface vectors, the second column the number of
 pentacene molecules in the cell. $\varepsilon_{11}$,
  $\varepsilon_{22}$, $\varepsilon_{12}$ are the components of the
  strain tensor applied to the PC surface cell in order to match the gold supercell. $\bar{\varepsilon} =(|\varepsilon_{11}|+
  |\varepsilon_{22}|+ |\varepsilon_{12}|)/3$, is the average strain.}
\begin{ruledtabular}
\begin{tabular}{lccccc}
  Au(111) & PC(010)  & $\varepsilon_{11}$  & $\varepsilon_{22}$  &
  $\varepsilon_{12}$  & $\bar{\varepsilon}$ \\
      \cline{1-6}
$(\sqrt{3}\times 6  )$ &  1  &  -16.0 & 10.9 & 3.3  &    10.1 \\
$(2\times 3\sqrt{3} )$ & 1 &  -3.0  & -4.0 &  2.9 &    3.3  \\
$(2\times 3\sqrt{7} )$ &  1  &  -6.4  & -0.54  &  0.0  &   2.3  \\
$(2\times \sqrt{199} )$ & 2  &   0.2  &  0.7  &  3.7  &  1.5  \\
$(\sqrt{43} \times 7 )$  & 4 & -0.8  &  -0.3 &  -0.5 & 0.5   \\
$(\sqrt{103} \times \sqrt{273} )$  & 12 & 0.0  &  0.1 &  0.1 & 0.1   \\
\end{tabular}
\end{ruledtabular}
\end{table}

\subsection{Adsorption geometries}

In the following we will calculate the relaxed geometry of pentacene on the
Au(111)-$(\sqrt{3}\times 6)$, Au(111)-$(2\times 3\sqrt{3})$, and Au(111)-$
(2\times 3\sqrt{7})$ surfaces and compare the adsorption energies. In each
case the starting geometry is obtained by matching the PC crystal with the
corresponding Au lattice, setting the initial Au-pentacene distance to 3.1\AA
. Other details of the relaxation procedure are described in section~\ref{sec:methodology}.

\begin{figure}[tbp]
\includegraphics[width=\linewidth]{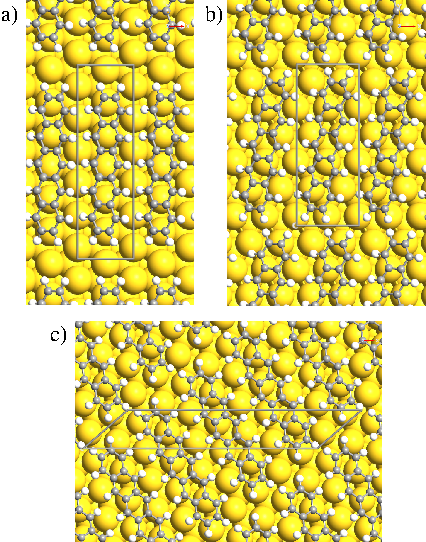}

\caption{ Top view of the a)
    $(\sqrt{3}\times6)$,  b) $(2 \times 3 \sqrt{3})$, and c) $(2
    \times 3 \sqrt{7})$ geometry.
     }  \label{fig:configurations}
\end{figure}

\begin{figure}[tbp]
\includegraphics[width=\linewidth]{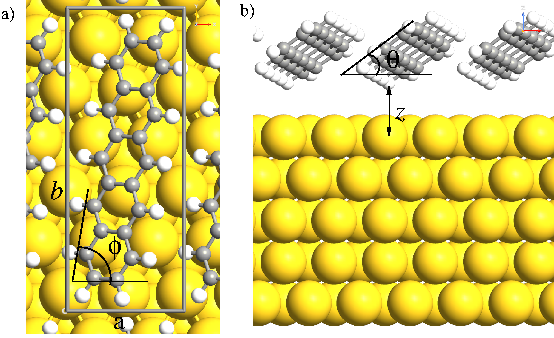}

\caption{ a) Top view  $(2 \times 3 \sqrt{3})$ geometry illustrating
    the pentacene surface lattice parameters $a$, $b$ and the angle
    between the pentacene long axis and the a vector, $\phi$. b) Side view
    illustrating the adsorption height, $z$, and the angle of the
    molecular plane to the gold surface, $\theta$.
     }  \label{fig:symbols}
\end{figure}

The geometries of the relaxed configurations are illustrated in Fig.~\ref{fig:configurations}. For each relaxed geometry we calculate the pentacene
lattice vectors $a, b$, adsorption height $z$ and adsorption angles $\theta,
\phi$, the definitions of the parameters are illustrated for geometry
Au(111)-$(2\times 3\sqrt{3} )$ in Fig.~\ref{fig:symbols}. The main results
of the geometry optimizations are summarized in Table~\ref{pentacene-au111}
together with available experimental data.

\begin{table}
\caption{\label{pentacene-au111}
Data for the relaxed geometries of pentacene on the Au(111) surface, for
the $(\sqrt{3} \times 6)$, $(2\times 3\sqrt{3} )$ and $(2\times
3\sqrt{7} )$ supercell. For definition of geometry parameters $a$,
$b$,  $z$, $\theta$ and $\phi$, see Fig.~\ref{fig:symbols}. $E_c$ is the binding energy of the
pentacene on gold. $\Phi_{Au}$  is the workfunction of the clean gold surface,
$\Phi$ is the work function of the surface with the adsorbed
pentacene, and
$\Delta \Phi$ is the relative change in the surface work function. Reference
calculation values  for the $(\sqrt{3} \times 6)$ surface from Li
{\it et. al.}\cite{Li-2009, note-z} are given in the third column.
}
\begin{ruledtabular}
\begin{tabular}{cccccc}
 & $(\sqrt{3} \times 6)$ & Li {\it et. al.}\cite{Li-2009} & $(2 \times
  3\sqrt{3})$  & $(2 \times 3\sqrt{7})$ & Exp.\\ \hline
      \cline{1-6}
$a$ (\AA) &  5.11   &5.11  &  5.90  & 5.90 & 5.64\cite{Schroeder-2002}
      5.76\cite{France-2003} 5.7\cite{Kafer-2007, Kang-2003} \\
$b$ (\AA)&  17.71  & 17.71  &  15.44  & 15.33 & 14.8\cite{Schroeder-2002}
      15.0\cite{France-2003} 15.5\cite{Kafer-2007, Kang-2003}\\
$z$ (\AA) &  3.35  & 3.1-3.5  &  3.18  & 3.17\\
$\theta$ & 40$^0$ & 38$^0$ & 36$^0$ & 34$^0$ & 43$^0$\cite{Ihm-2006} 31$^0$\cite{Kafer-2007}  \\
$\phi$ & 87$^0$ & 81$^0$ & 81$^0$ & 80$^0$ \\
$E_c$ (eV) & -0.29  & -0.16 & -0.42 & -0.42 & -1.14\cite{France-2003}  \\
$\Phi_{Au}$  (eV) & 5.19  & 5.25  & 5.19  & 5.16 &  5.47 \cite{Schroeder-2002}
5.4\cite{Watkins-2002} 5.1\cite{Diao-2007}\\
$\Phi$  (eV) &   4.25  &  4.29  & 4.48 & 4.50 & 4.52 \cite{Schroeder-2002}
4.4\cite{Watkins-2002} 4.6\cite{Diao-2007}\\
$\Delta \Phi$  (eV) & -0.94  & -0.96  & -0.71 & -0.66 &
-0.95\cite{Schroeder-2002,France-2003} -1.0\cite{Watkins-2002} -0.5\cite{Diao-2007}\\
\end{tabular}
\end{ruledtabular}
\end{table}

We first discuss the comparison of the calculated values for the  $(\sqrt{3}
\times 6)$ with Li \textit{et. al.}\cite{Li-2009}. The latter was  done
using a plane-wave method, thus, discrepancy with this  calculation gives a
good estimate of the accuracy of our method. The geometry parameters, $a$, $b
$, $z$, $\theta$ and $\phi$ are very similar for the two methods\cite{note-z}. In the current work we use a relaxation threshold of 0.01 eV, while Li
\textit{et. al.}\cite{Li-2009} use 0.03 eV, this may explain the small
discrepancy between the calculations.

The adsorption energy is 0.16 eV higher than the value of Li \textit{et. al.}%
\cite{Li-2009}. This is a rather large deviation, in particular since in the
study by Lee \textit{et. al.} they found that a LCAO method with long range
orbitals applying the BSSE correction should have an accuracy of 0.02 eV
compared with a plane-wave calculation. The adsorption energy is however in
good agreement with the work by Lee \textit{et. al.}\cite{Lee-2007} where
they found a pentacene-gold interaction energy of -0.28 eV for an isolated
pentacene molecule adsorbed on the Au(100) surface\cite{Lee-2007} using the
GGA functional of Perdew-Burke-Ernzerhof\cite{Perdew-1996-PRL}. In the
present study there are also interactions within the pentacene molecular
layer which should give an additional adsorption energy, in accordance with
that our value  is slightly higher than the value by Lee \textit{et. al.}.
Thus, we are confident in our calculated adsorption energy, and believe that
the value reported by Li \textit{et. al.}\cite{Li-2009} is too low.

We also calculated the change in the Au(111) work function upon adsorption
of pentacene. We find a reduction by -0.94 eV, in excellent agreement with
the -0.96 eV found by Li \textit{et. al.}\cite{Li-2009}.

The adsorption geometry and properties of the $(2\times 3\sqrt{3})$ and $(2\times 3\sqrt{7})$ structure are almost identical and in the following we
will focus on the $(2\times 3\sqrt{3})$ structure. The adsorption energy in
this structure is significantly higher than in the $(\sqrt{3}\times 6)$
structure. It is interesting to divide the adsorption energy into
pentacene-pentacene interactions, $E^{\mathrm{p-p}}$, and pentacene-Au(111)
interaction, $E^{\mathrm{au-p}}$. This separation of the energy can be
obtained by performing a calculation of the pentacene overlayer without the
gold surface. For the $(\sqrt{3}\times 6)$ structure we find $E^{\mathrm{p-p}}=-0.22$ eV and $E^{\mathrm{au-p}}=-0.07$ eV, while for the $(2\times 3\sqrt{3})$ structure $E^{\mathrm{p-p}}=-0.30$ eV and $E^{\mathrm{au-p}}=-0.12$ eV.
Thus, the main difference in the adsorption energy between the two
structures arises from a stronger overlayer binding energy in the $(2\times 3
\sqrt{3})$ structure.

The workfunction change upon adsorption of pentacene is -0.71 eV for the $(2\times 3 \sqrt{3})$ geometry, which is significantly lower than what was
calculated for the $(\sqrt{3}\times6)$ surface. Table~\ref{pentacene-au111}
shows experimental data for the work function change upon adsorption of
pentacene. Note the wide range of values [-1.0 eV, -0.5 eV]. Inspecting the
numbers show that the wide spread arises from a spread in the value for the
work function of the clean gold surface. The values of the work function for
the pentacene covered gold surface seem more reliable and are in the the
range 4.4-4.6 eV, in excellent agreement with our value for the $(2 \times 3
\sqrt{3})$ structure.

\begin{figure}[tbp]
\includegraphics[width=\linewidth]{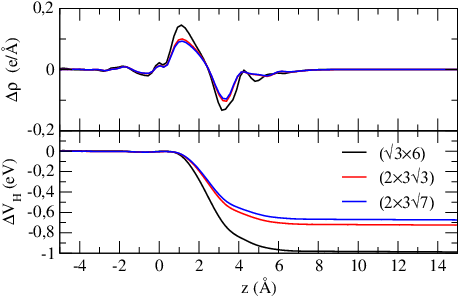}

\caption{ The induced density and electro-static potential upon
    adsorption of pentacene on the Au(111) surface for the
    $(\sqrt{3}\times6)$ (black),  $(2 \times 3
    \sqrt{3})$(red), and  $(2 \times 3  \sqrt{7})$ geometry (blue). $z=0$ is the position of the
    top most gold atoms.
     }  \label{fig:vn}
\end{figure}

Fig.~\ref{fig:vn} shows the electron density for the combined
Au(111)-pentacene system, subtracted the electron density of the separated
systems. The result for the $(\sqrt{3}\times6)$ cell is in excellent
agreement with Li \textit{et. al.}\cite{Li-2009}. The main change in the
density is between the gold surface and the pentacene molecule, and the
molecule pushes the density back towards the surface, known as the pillow
effect.\cite{Koch-2007b} The density change for the $(2 \times 3 \sqrt{3})$
surface is qualitatively the same, however, the change in density is
slightly smaller and the change in electro-static potential therefore
correspondingly lower.

\begin{figure}[tbp]
\includegraphics[width=\linewidth]{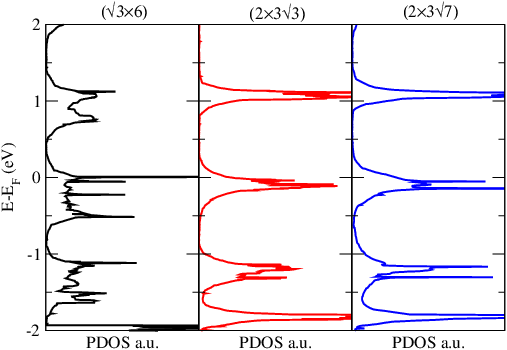}

\caption{ The Projected Density Of States (PDOS) of the pentacene
    molecule in the     $(\sqrt{3}\times6)$ (black),  $(2 \times 3
    \sqrt{3})$(red), and  $(2 \times 3  \sqrt{7})$ geometry (blue).
     }  \label{fig:pdos}
\end{figure}

To further investigate the difference between the two systems Fig.~\ref{fig:pdos} shows the Projected Density Of States (PDOS) of the pentacene
layer for the 3 systems. We see that the peaks are much broader for the $(\sqrt{3}\times6)$ system. We relate this to a stronger pentacene-pentacene
interaction along the $b$ direction, due to the 15 \% smaller lattice
constant in this direction compared to the $(2 \times 3  \sqrt{3})$ and $(2
\times 3 \sqrt{7})$ systems.

We also note that the HOMO is located at the gold fermi level, thus the
pentacene crystal will be hole doped by the gold surface. The HOMO-LUMO gap
is ~1.2 eV, significantly smaller than the 2.2 eV found experimentally for
the Au(100) surface\cite{McDonald-2006}, which is related to the inability
of PW91 to accurately describe the unoccupied states, thus, leading to a too
small band gap.

\section{Conclusions}

\label{sec:conclusions} We have investigated the geometry of a single layer
of pentacene on the Au(111) surface. We presented a new method for finding
the preferred surface lattice of the system by strain matching the pentacene
crystal with the Au(111) surface. With this approach we find the $(2 \times
3 \sqrt{3})$ structure to have a low strain, and we find that the adsorption
energy of this structure is more favourable than the $(\sqrt{3}\times6)$
structure suggested in a previous study\cite{Li-2009}. This structure is in
agreement with experimental data for the pentacene surface geometry\cite{Koch-2007} and we reproduce the measured workfunction of the pentacene
covered Au(111) \cite{Schroeder-2002, Watkins-2002,Diao-2007}. We suggest
that the presented method may be used for determining the adsorption
structure of other weakly adsorbed molecular layers.

\begin{acknowledgments}
\label{sec:acknowledgements} We acknowledge Alexander Bratkovski for proof
reading the manuscript.
\end{acknowledgments}

\bibliography{paper}

\end{document}